\documentclass[english,twocolumn,showkeys,showpacs,preprintnumbers,amsmath,amssymb,prl]{revtex4}
\usepackage[T1]{fontenc}
\usepackage[latin1]{inputenc}
\usepackage{prettyref}
\usepackage{amsmath}
\usepackage{color}

\makeatletter
\usepackage{lmodern}



\usepackage{babel}
\makeatother
\begin{document}

\title{Isoconfigurational thermostat}

\author{Igor Rychkov}

\author{Debra J. Searles}

\email{D.Bernhardt@griffith.edu.au}

\affiliation{Nanoscale Science and Technology Centre, School of Biomolecular and
Physical Sciences, Griffth University, Brisbane, Qld 4111 Australia}

\date{\today}

\begin{abstract}
A holonomic constraint is used to enforce a constant instantaneous
configurational temperature on an equilibrium system. Three sets of
equations of motion are obtained, differing according to the way in
which the holonomic constraint is introduced and the phase space distribution
function that is preserved by the dynamics. Firstly, Gauss' principle
of least constraint is used, and it is shown that it does not preserve
the canonical distribution. Secondly, a modified Hamiltonian is used
to find a dynamics that provides a restricted microcanonical distribution.
Lastly, we provide equations that are specifically designed to both
satisfy the temperature constraint and produce a restricted canonical
ensemble. 
\end{abstract}

\pacs{05.10.-a, 05.20.-y, 05.20.G}

\keywords{statistical physics, temperature, molecular dynamics}

\maketitle
When simulating a molecular system at equilibrium one would often
like to have it at a constant temperature rather than the energy.
In a non-equilibrium simulation controlling the temperature becomes
mandatory as the driving forces would otherwise heat up the sytems
indefinitely. In order to thermostat a system in bulk, away from the
walls, some synthetic fields can be added to the equations of motion.
Two important questions arise. Firstly, what microscipic measure should
one choose for the temperature to be controlled? Secondly, what probability
distribution will the resulting equations generate in the system phase
space.

One popular thermostat widely used in molecular dynamics simulations
is the isokinetic Gaussian equations of motion, which fix the kinetic
energy to a desired value at all times and generate the isokinetic
canonical distribution \citep{Evans1990}. Unfortunately, although
the kinetic energy of molecules is a good measure of the temperature,
the thermal velocities cannot always be distinguished from the streaming
velocity which is usually unknown beforehand. If the streaming velocity
profile is presumed, the thermostat will maintain it, acting as a
{}``profile-biased'' thermostat (PBT), which is often undesirable
\citep{Delhommelle2003}. 

There are configurational measures for the temperature which do not
require the peculiar, thermal velocities to be known. Several expressions,
all equal to one another in the thermodynamics limit, are known \citep{Owen2000};
the most popular one being (in various notations)\begin{multline}
T=T_{conF}\equiv\frac{\left\langle \nabla U\cdot\nabla U\right\rangle }{\left\langle \nabla\cdot\nabla U\right\rangle }=\\
\frac{\left\langle \frac{\partial U}{\partial q}\cdot\frac{\partial U}{\partial q}\right\rangle }{\left\langle \frac{\partial}{\partial q}\cdot\frac{\partial U}{\partial q}\right\rangle }=\frac{\sum_{i}^{Nd}\left\langle \frac{\partial U}{\partial q^{i}}\frac{\partial U}{\partial q^{i}}\right\rangle }{\sum_{i}^{Nd}\left\langle \frac{\partial}{\partial q^{i}}\frac{\partial U}{\partial q^{i}}\right\rangle }\label{eq:Tconf}\end{multline}
where $T_{conF}$ is the configurational temperature in energetic
units, $U$ is the interaction potential energy, $q$ is a vector
containing the position of all the particles, $d$ is the Cartesian
dimension of the system, $N$ is the number of particles, $i$ labels
the degrees of freedom and the angular brackets represent an ensemble
average. This expression has been known as the hypervirial theorem
\citep{Gray1984} and has been proved for canonical and microcanonical
ensembles \citep{Landau1980,Owen2000,Rugh1998}.

The first successful implementation of a thermostat controlling the
configuational temperature was with the Nosé-Hoover method. The spurious
string phase observed in shear flow simualtions with PBT was eliminated
when using the configuational thermostat \citep{Delhommelle2003}.
Moreover, in the most recent revision of the method \citep{Braga2005}
the projected phase space distribution was made canonical.

In the Nosé-Hoover method the dynamics is extended by a degree of
freedom that describes coupling to the thermostat, and the fluctuations
of the temperature are governed by an extra parameter. It is of interest
to see if a practical, constant configurational temperature thermostat
can be developed that constrains the original system to a constant
\emph{instantaneous} configurational temperature. We take Eq.\eqref{eq:Tconf}
without the averaging brackets as a measure of instantaneous configurational
temperature, by analogy to the Gaussian isokinetic thermostat which
constrains the instantaneous kinetic energy. In this Letter we consider
different ways of introducing the constraint into the original equilibrium
Hamiltonian equations and the effect on the ensemble probability distribution.
In fact, the resulting equation will be valid not only for the configurational
thermostat but for any holonomic constraint.

A constraint equation is insufficient to determine the constraint
reaction forces: an additional principle has to be employed in order
to find the constrained equations of motion \citep{Jose1998}. One
can use principles restricting the reaction forces, for instance,
d'Alambert's principle of orthogonal forces or Gauss' principle of
least constraint. Alternatively, there is also the Hamilton variational
principle of least action. For holonomic constraints, they all lead
to the same equations of motion: \textcolor{black}{\begin{equation}
\ddot{q}=-\nabla U-\lambda(q,\dot{q})n=-(\nabla U)_{||}-n\left(\dot{q}\cdot\frac{\nabla g}{|\nabla h|}\right)\label{eq:Lagrangian acceleration}\end{equation}
where $q$ is a position vector of all the particles,} $\lambda(q,\dot{q})=\dot{q}\cdot|\nabla h|^{-1}\nabla g-n\cdot\nabla U$
\textcolor{black}{is a Lagrange multiplier determined as to satisfy
the equation of constraint $h(q)=h_{0}$; $g(q,\dot{q})\equiv\dot{h}=\nabla h\cdot\dot{q}$
is the conjugate non-holonomic constraint; $n=\tfrac{\nabla h}{|\nabla h|}$
is the unit normal vector to the constraint hypersurface $\Omega_{h}$,}
and $(\nabla U)_{||}$ is the component of $\nabla U$ tangent to
\textcolor{black}{$\Omega_{h}$}. In the following we assume unit
mass, for simplicity.

One can move to the Hamiltonian formalism by introducing the generalized
momenta simply as $p=\dot{q}$. This, together with \eqref{eq:Lagrangian acceleration}
give the standard Gaussian equations of motion with an isoconfigurational
thermostat. The same equations are obtained through Dirac's general
treatment of constrained Hamiltonian systems \citep{Dirac1950} when
the constraint is holonomic:\begin{equation}
\left\{ \begin{aligned}\dot{q} & =p\\
\dot{p} & =-\nabla U-\lambda(q,p)n=(\nabla U)_{||}-n\left(p\cdot\frac{\nabla g}{|\nabla h|}\right)\end{aligned}
\right.\label{eq:dynamics1}\end{equation}
Let us consider the phase space properties of this dynamical system.
Besides the constrained quantity itself, the energy of the system
$E(q,p)\equiv\frac{1}{2}p\cdot p+U$ is also a constant along the
actual trajectory: \begin{equation}
\dot{E}(q,p)=\frac{\partial E}{\partial q}\cdot\dot{q}+\frac{\partial E}{\partial p}\cdot\dot{p}=0-\lambda(q,p)(p\cdot n)=0\label{eq:dH/dt}\end{equation}
 Moreover, if the constraint depends only on intramolecular distances,
as the configurational temperature does, then the total momentum $P$
is preserved as well. We now investigate the evolution of the phase
space elementary volumes by calculating the phase space compression
factor:\begin{multline}
\Lambda\equiv\frac{\partial}{\partial q}\cdot\dot{q}+\frac{\partial}{\partial p}\cdot\dot{p}=0-n\cdot\frac{\partial\lambda}{\partial p}=-2n\cdot\frac{\nabla g}{|\nabla h|}=\\
-2\frac{\nabla h\cdot\nabla\dot{h}}{\nabla h\cdot\nabla h}=-2\frac{\nabla h\cdot\frac{d}{dt}\nabla h}{\nabla h\cdot\nabla h}=-\frac{d}{dt}\ln(\nabla h\cdot\nabla h)\label{eq:Lambda}\end{multline}
 If we write the Liouville equation with the phase space compression
factor \citep{Evans1990}, \begin{equation}
\frac{d\ln f}{dt}=-\Lambda\label{eq:Liouville}\end{equation}
the probability distribution density function $f$ is easily found
and is written over the whole phase space as follows,\begin{equation}
f=\nabla h\cdot\nabla h\,\delta(h-h_{0})\delta(g)\delta(E-E_{0})\delta(P-P_{0})\label{eq:f1}\end{equation}

For some constraints the weight $\nabla h\cdot\nabla h$ amounts to
a constant. This is the case for the bond length constraints, for
instance. Generally, however, the distribution \prettyref{eq:f1}
is non-uniform over equal phase space volumes. However, the volumes
themselves are constantly being stretched and compressed by the dynamics
as the phase space compression factor is non-zero \prettyref{eq:Lambda}
(albeit being zero on average). If one uses the so called invariant
measure of the phase space \citep{Tuckerman1999}, which for the dynamics
\prettyref{eq:dynamics1}turns out to be just the $\nabla h\cdot\nabla h\, dqdp$
\citep{Melchionna2000}, then the density function becomes uniform
over non-equal but invariant volumes. Whichever the interpretation
of this weight, the dynamics will only preserve a non-canonical distribution
unless $\nabla h\cdot\nabla h$ is constant, and the values accumulated
during the simulation would have to be {}``reweighted'' before any
comparison with experimental data is possible. It is preferable to
find such isoconfigurational thermostats that would produce a microcanonical,
or better still, a portion of the Gibbs canonical distribution --
a restricted canonical distribution.

In order for the configuration to satisfy a holonomic constraint,
the velocities should be tangent to the constraint hypersurface: $\dot{q}\cdot n=0$.
Let us take an arbitrary vector $p$ that we can call a peculiar momentum
and subtract from it its component, which we name a {}``convection''
term, perpendicular to the hypersurface: $\dot{q}=p-n(p\cdot n)$.
We now use the internal energy expressed in the coordinates $(q,p)$,
\begin{equation}
H=\frac{\dot{q}\cdot\dot{q}}{2}+U=\frac{p\cdot p-(p\cdot n)^{2}}{2}+U\label{eq:ArnoldH}\end{equation}
as a Hamiltonian to generate the following equations of motion:\begin{equation}
\left\{ \begin{aligned}\dot{q} & =p-n(p\cdot n)\\
\dot{p} & =-\nabla U+(p\cdot n)\nabla(p\cdot n)\end{aligned}
\right.\label{eq:dynamics2}\end{equation}
The Hamiltonian \prettyref{eq:ArnoldH} is known to be Hamiltonian
in redundant coordinates \citep{ArnoldCelestial}. As for any other
Hamiltonian system, the phase space compression factor is zero, and
the distribution is the (restricted) microcanonical,\begin{equation}
f_{\mu}=\delta(h-h_{0})\delta(g)\delta(H-H_{0})\delta(P-P_{0}).\label{eq:f2}\end{equation}

Finally, we consider a more general family of equations satisfying
the holonomic constraint:\begin{equation}
\left\{ \begin{aligned}\dot{q} & =p-n(p\cdot n)\\
\dot{p} & =-(\nabla U)_{||}+R(q)\end{aligned}
\right.\label{eq:dynamics3}\end{equation}
where we excluded from the unknown reaction force $R$ a term that
is to cancel the normal component of the intermolecular interaction
force and assumed that $R$ should only depend on the configurational
information and not on the momenta. We will demand the restricted
canonical distribution,\begin{equation}
f_{c}=e^{-\beta H}\delta(h-h_{0})\delta(g)\delta(P-P_{0})\label{eq:Gibbs}\end{equation}
where $\beta=1/T_{0}$. According to the Liouville equation \eqref{eq:Liouville},
we must then have \[
\Lambda=-\frac{d}{dt}\ln f=\beta\dot{H}\]
As we calculate the $\Lambda$ and $\dot{H}$ similarly to \eqref{eq:Lambda}
and \eqref{eq:dH/dt} for the system \eqref{eq:dynamics3}, we find
the following relation for $R$:\[
-\nabla\cdot n(p\cdot n)=\beta p\cdot R\]
where the left-hand side can be further modified to see the scalar
product of $p$,\[
-\nabla\cdot n(p\cdot n)=p\cdot(\nabla\cdot n)n\]
Since it must be valid for any $p$ we obtain the solution,\begin{equation}
R=-T_{0}(\nabla\cdot n)n=-T_{0}[n(\nabla\cdot n)+(n\cdot\nabla)n]\label{eq:R}\end{equation}
where the first term in the brackets is normal and the second is tangential
to the $\Omega_{h}$ (because $n\cdot(n\cdot\nabla)n=\frac{1}{2}(n\cdot\nabla)(n\cdot n)=0$).

When $T_{conF}=T_{0}$ is chosen as the holonomic constraint $h(q)=h_{0}$,
the equations \eqref{eq:dynamics3} and \eqref{eq:R} describe a dynamics
that preserves the configurational temperature at each time step and
generates a canonical distribution \eqref{eq:Gibbs} on the constraint
hypersurface. 

In conclusion, three sets of dynamic equation were derived describing
three different isoconfigurational ensembles: non-canonical \eqref{eq:dynamics1},
\eqref{eq:f1}; microcanonical \eqref{eq:dynamics2}, \eqref{eq:f2};
and canonical \eqref{eq:dynamics3}, \eqref{eq:Gibbs}. The canonical
isoconfigurational ensemble is deemed to be the best candidate to
simulate a system at a given temperature that is established instantaneously
throughout the system based only on the configurational information,
which should be especially useful in simulation of flow when the streaming
velocity profile is unknown. Work is underway by the authors to numerically
test the equations obtained and specifically resolve the implementation
difficulties, such the algebraic unwieldiness and numerical stiffness
of higher derivatives of potential energy and the choice of the starting
configuration that corresponds to the desired temperature.

\bibliographystyle{apsrev}
\bibliography{Bibliography/Configuational_Thermostat}

\end{document}